\documentclass[a4paper,10pt,twoside]{cpc-hepnp}

\usepackage{multicol}
\usepackage{graphicx}
\usepackage{booktabs}
\usepackage{amssymb,bm,mathrsfs,bbm,amscd,color}
\usepackage[tbtags]{amsmath}
\usepackage{lastpage}
\usepackage{CJK}

\begin{document}
\begin{CJK*}{GBK}{song}

\fancyhead[oc]{\small QIAN Xiang-Li et al.: Calibration of pulse transit time through a cable for EAS experiments}

\title{Calibration of pulse transit time through a cable for EAS experiments\thanks{Supported by the Ministry of Science and Technology of China, the Natural Sciences Foundation of China (Nos. 11135010, 11105156), by the 973 Program of China(No. 2013CB837000), and by Youth Innovation Promotion Association, Chinese Academy of Sciences. }}
\author{%
QIAN Xiang-Li$^{1,2}$\email{zhangyi@mail.ihep.ac.cn}%
\quad CHANG Jin-Fan$^{1}$\quad FENG Cun-Feng$^{2}$\quad FENG Zhao-Yang$^{1}$\\
\quad GOU Quan-Bu$^{1}$\quad GUO Yi-Qing$^{1}$\quad HU Hong-Bo$^{1}$\quad LIU Cheng$^{1}$\\
\quad WANG Zheng$^{1}$\quad XUE Liang$^{2}$\quad ZHANG Xue-Yao$^{2}$\quad ZHANG Yi$^{1;1)}$
}
%\author{%
%QIAN Xiang-Li(钱祥利)$^{1,2}$\email{zhangyi@mail.ihep.ac.cn}%
%\quad CHANG Jin-Fan(常劲帆)$^{1}$\quad FENG Cun-Feng(冯存峰)$^{2}$\quad FENG Zhao-Yang(冯朝阳)$^{1}$\\
%\quad GOU Quan-Bu(苟全补)$^{1}$\quad GUO Yi-Qing(郭义庆)$^{1}$\quad HU Hong-Bo(胡红波)$^{1}$\quad LIU %Cheng(刘成)$^{1}$\\
%\quad WANG Zheng(王铮)$^{1}$\quad XUE Liang(薛良)$^{2}$\quad ZHANG Xue-Yao(张学尧)$^{2}$\quad ZHANG %Yi(张毅)$^{1;1)}$
%}
\maketitle

\address{%
$^1$ Institute of High Energy Physics, Chinese Academy of Sciences, Beijing 100049, China\\
$^2$ Shandong University, Jinan 250100, China\\
}

\begin{abstract}
In ground-based extensive air shower experiments, the direction and energy are reconstructed by measuring the relative arrival time of secondary particles, and the energy they deposit. The measurement precision of the arrival time is crucial for determination of the angular resolution. For this purpose, we need to obtain a precise relative time offset for each detector, and to apply the calibration process. The time offset is associated with the photomultiplier tube, cable, relevant electronic circuits, etc. In view of the transit time through long cables being heavily dependent on the ambient temperature, a real-time calibration method for the cable transit time is investigated in this paper. Even with a poor-resolution time-to-digital converter, this method can achieve high precision. This has been successfully demonstrated with the Front-End-Electronic board used in the Daya Bay neutrino experiment.
\end{abstract}
%地基广延大气簇射实验，通过探测次级粒子的相对到达时间和沉积能量来重建原初宇宙线的方向和能量。
%方向重建的精度由次级粒子到达时间的测量精度决定。为了提高测量精度，需要精确获得每个探测器的
%相对时间偏移量，因此时间标定是必不可少的。探测器的时间偏移量与光电倍增管、电缆、相关电子学
%等有关。其中，脉冲信号在电缆中的传输时间受环境温度影响很大。本文研究了一种标定脉冲在长电缆
%中的传输时间的方法。该方法已用大亚湾中微子实验的前端电子学板做了成功验证。即使用精度不高的
%时间数字转换器，该方法也能得到很高的精度。

%广延大气簇射实验；时间游离效应；实时标定；传输时间

\footnotetext[0]{\hspace*{-3mm}\raisebox{0.3ex}{$\scriptstyle\copyright$}2013
Chinese Physical Society and the Institute of High Energy Physics
of the Chinese Academy of Sciences and the Institute
of Modern Physics of the Chinese Academy of Sciences and IOP Publishing Ltd}%

\begin{multicols}{2}

\section{Introduction}
An extensive air shower (EAS) is a broad cascade of ionized particles and electromagnetic radiation produced when an energetic primary cosmic ray (CR) enters the atmosphere. EAS arrays directly detect the secondary particles that reach observational altitudes. Typical EAS detectors have a wide field-of-view of 2sr and operate nearly 100\% of the duty cycle. These two characteristics make them ideal instruments for studying cosmic ray anisotropy\cite{lab1}, very high energy (VHE) gamma rays\cite{lab2}, extended sources\cite{lab3} and transient phenomena\cite{lab4}. In recent years, the Tibet AS$\gamma$\cite{lab5}, Milagro\cite{lab6} and ARGO-YBJ\cite{lab7} experiments have made significant progress in TeV $\gamma$-ray astronomy, CR anisotropy and knee physics. The next generation EAS experiments, HAWC\cite{lab8} and the LHAASO project\cite{lab9}, are expected to answer these long-standing problems in astrophysics, and hence to make more contributions to the CR community.

The direction of the primary CR is reconstructed according to the relative arrival time of secondary particles as they hit each detector of the array. The arrival time and the deposited energy of the secondary particles are recorded by the detector as the hit time and the deposited charge, respectively. Generally, particles within several tens of meters from the shower core are expected to form a curved front. By fitting the air shower front using the hit time and the deposited charge, the direction and energy of the primary CR are reconstructed. The measurement precision of the arrival time is crucial for determining the direction of the primary cosmic ray. In order to obtain a good angular resolution, it is essential to precisely calibrate the relative time offset of each detector.

Considering a scintillation detector as an example, the time offset of the detector is related to the scintillator, light guide, photomultiplier tube (PMT), cable, relevant electronic circuits, and so on. When the secondary particles hit the scintillator, the atoms (molecules) are ionized and photons are generated in the deexcitation process. The emission times of these photons follow an exponential distribution. The photons then propagate through the light guide box, the geometry of which is closely related to the propagation delay. Photoelectrons are emitted as soon as the photons hit the photocathode, and these electrons are multiplied from the first dynode to the last dynode. The transit time spread is different in different models of PMT. Finally, the pulse formed by these electrons propagates through the long cable and the electronic circuits. The charge and arrival time of the pulse are digitized by time-to-digital converters (TDC) and analog-to-digital converters (ADC), respectively. In general, most of the parameters are constant, and need to be calibrated only once for a long period of time (usually one year). Only those parameters strongly dependent on the ambient temperature and atmospheric pressure are monitored in real-time calibration.

In EAS experiments with a large area, long cables are always used to transfer the analog signal from the detector to the data acquisition (DAQ) system. Most of the cables are placed outside along the cable trenches. However, the transit time of the pulse through the cable is greatly affected by the ambient temperature. For example, the Tibet air shower array (AS) consists of 761 FT counters with a spacing of 7.5 m, covering an area of 36900 $m^{2}$ surrounded by 28 D counters\cite{lab10}. All the signals from the FT counters are connected to the DAQ system by 150 m long cables. Considering that the Tibet AS array is located at an altitude of 4.3 km, the temperature varies greatly over a single day, e.g. from $-25^{0}C$ to $15^{0}C$ in winter. The maximum variation in the transit time of the cables can reach $\sim$ 3 ns; in this case the typical temperature coefficient of the cables is about 0.01\% per degree. Generally, a precision of 0.1 ns is needed in calibrating the pulse transit time in EAS experiments.

In this paper, an improved calibration method for the pulse transit time through the cable is proposed and the desired precision of 0.1 ns is obtained even in case of poor TDC resolution. This method and precision have been successfully demonstrated with the Front-End-Electronic (FEE) board used in the Daya Bay experiment\cite{lab11}.

\section{TDR method for measuring the pulse transit time through a cable}
The most popular method to measure cable length, and thus transit time, is time domain reflectometry (TDR) analysis\cite{lab12,lab13}. This technique comes from the development of high-speed pulse technology and is widely used in measurement systems due to its non-destructive and high-precision features. The principle of TDR is described below.

If a pulse is emitted at one end of a cable, all or a part of the pulse will be reflected when the impedance of the other end is mismatched. The velocity of the pulse can be assumed to be constant. Thus, the cable length can be calculated by the formula
\begin{equation}\label{eq1}
  l=\frac{v\Delta T}{2},
\end{equation}
where $l$ is the cable length, $v$ is the velocity of the pulse transferred in the cable (determined by the type of cable), and $\Delta T$ is the time interval between the direct pulse and the reflected pulse.

The TDR method is restricted by two key factors in our case. Firstly, the high precision measurement of the time interval $\Delta T$ requires a high resolution TDC, i.e. the least significant bit (LSB). It is not necessary to choose an extremely high precision resolution TDC in EAS experiments, because the intrinsic time deviation of the shower front spreads to $\sim$ 1 ns. The second is the time-walk effect, i.e. the dependence of the threshold trigger time on the signal amplitude when using fast leading edge discriminators. Due to this effect, correctly identifying the arrival time of the reflected pulse is non-trivial.

\begin{center}
  \includegraphics[scale=0.47]{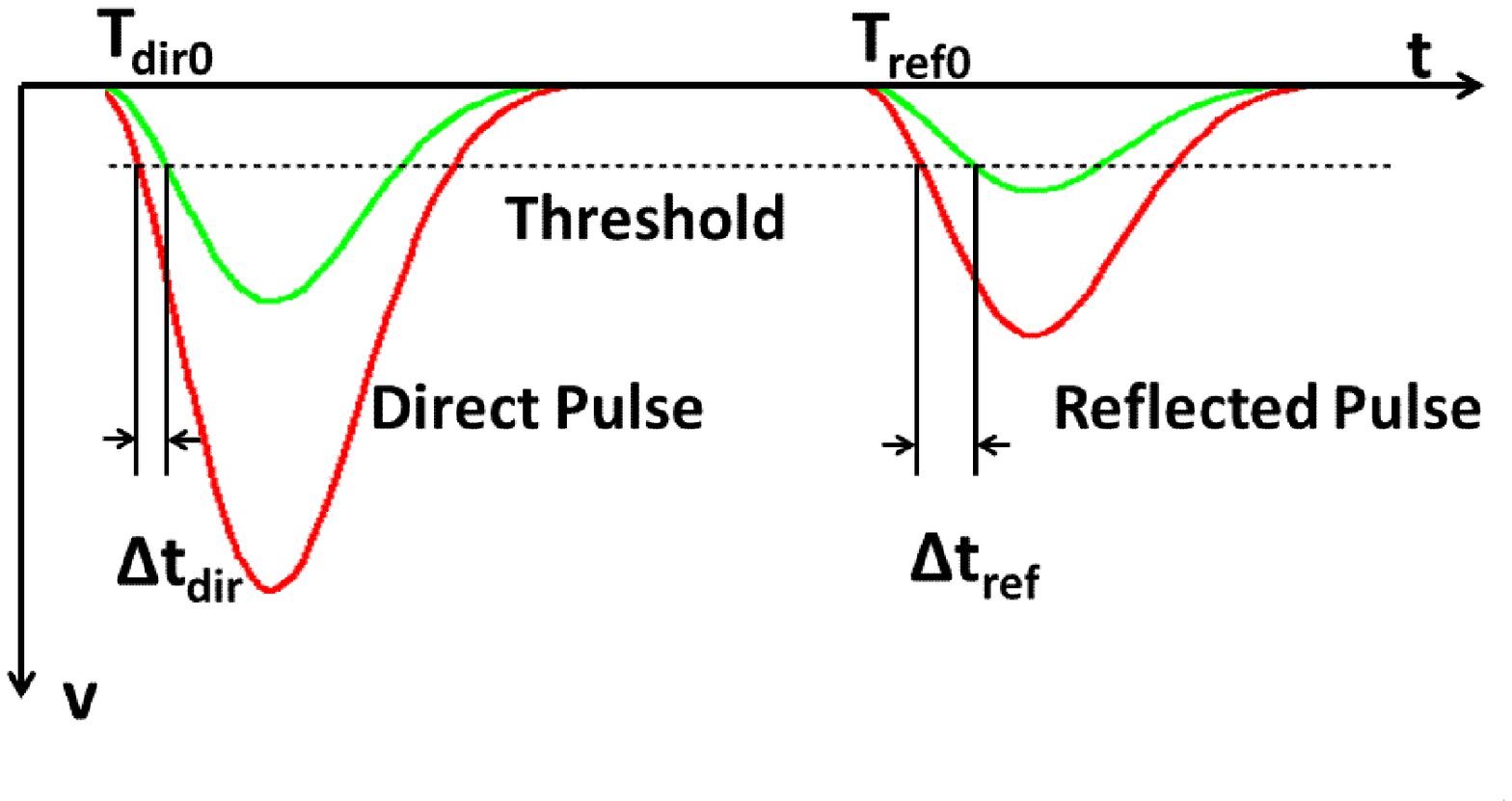}\\
  \figcaption{The time-walk effect for direct and reflected pulses. $T_{dir0}$ and $T_{ref0}$ are the start times of the two pulses. $\Delta t_{dir}$ and $\Delta t_{ref}$ indicate the time deviation caused by the time-walk effect.}\label{fig1}
\end{center}

The time-walk effect in TDR analysis is shown in Fig. \ref{fig1}, which is more significant for the reflected pulse. The reflected pulse has a lower height, due to attenuation when propagating twice through the cable. If the shape of the pulse is defined, the height is proportional to the charge of the pulse. The time-walk effect can be expressed by a relation of the start time and the charge of the pulse, and can be described by the formula \cite{timewalk2002,timewalk2007}
\begin{equation}\label{eq2}
  T_{th}(Q)=\frac{k}{Q^{m}}+T_{0},
\end{equation}
where $T_{th}(Q)$ is the start time measured by the TDC (i.e. the time at which the signal reaches the threshold), $Q$ is the charge of the pulse, $T_0$ is the real start time, and $m$ and $k$ are the time-walk factors, which are determined by the shape of the pulse.

For a rectangular-wave signal, the rise-time shape can be described as:
\begin{equation}\label{rshape}
  V(T)= {V_{\max }} \frac{{T - {T_0}}}{{{T_{Rise}}}},
\end{equation}
where $V_{max}$ is the maximum height of the pulse, $T_{Rise}$ is the rise time of the pulse and $V(T)$ is the pulse height at time T. The charge of the pulse should be proportional to the pulse height ($Q \propto {V_{\max }}$), thus the arrival time ($T_{th}$) at which the signal reaches the threshold ($V_{th}$) is
\begin{equation}\label{rshape2}
{T_{th}} - {T_0} \propto \frac{{{V_{th}}{T_{Rise}}}}{Q},
\end{equation}
which has the same form as Eq. (2). The corresponding time-walk factors are $m=1$ and $k \propto V_{th}T_{Rise}$.

If the transit time through the long cable is calibrated with a poor resolution TDC (e.g. 1.5 ns) using the TDR method, the calibrated resolution strongly depends on the TDC. In this case, the direct pulse can be produced with a constant amplitude and shape by the electronics, but the time-walk effect cannot be corrected for and results in a small time deviation for both direct and reflected pulses. The measured transit time will always be located in one bin of the TDC and the precision cannot be improved by increasing the statistics.

In this paper, an improved calibration method is proposed to solve this problem. The calibration signal is generated with different amplitudes but with a similar pulse shape. The time-walk effect is taken as a model by simultaneously measuring the charge and threshold triggering time of the direct and reflected pulses, respectively. The model as well as the measurement of the charge should in principle add more information for measuring the transit time.

\section{Real-time calibration method for pulse transit time through a cable}
A schematic diagram of the real-time calibration of cable transit time is shown in Fig. \ref{fig2}. The Daya Bay FEE board\cite{lab14} is used in this paper, as it is a candidate electronics board for the Tibet AS$\gamma$ experiment upgrade project. The Daya Bay FEE is a VME 9U module, which can handle up to 16 channels of PMT signals. In one module, 16 channels of dual range 12 bit ADC and 16 channels of 20 bit TDC are integrated. The waveform of each input pulse is be sampled with a 40 MHz clock after passing through a shaping circuit. The charge readout system is designed to digitalize charge in two ranges by the ADC and to preprocess data using a FPGA. The charge range is from 1.6 pC to 1800 pC, and the ADC resolution for the  fine range is 0.16 pC. In the time readout system, the FEE board adopts a time-control-delay method which is based on FPGA. The least significant bit (LSB) of the TDC is 1.5625 ns.

A calibration mode is also designed in the FEE board. A digital to analog converter (DAC) AD9726 provides calibration signals and sends them to the inputs of the FEE. The amplitude of the DAC signal is determined by a register which can be set through the VME bus. The DAC AD9726 is 16-bit, providing leading-edge performance at conversion rates of up to 400 million samples per second (MSPS). The differential nonlinearity is less than $\pm$0.5 LSB and the integral nonlinearity is less than $\pm$1.0 LSB. The register of the DAC can be set from 0 to 65535, representing $\sim$ 0-1834 pC of the direct pulse charge. In fact, the charge $Q$ recorded by the FEE board is proportional to the value of the DAC register. Using the DAC value instead of $Q$ will not affect the estimation of the parameter $T_0$ in Eq. (\ref{eq2}).
\begin{center}
\nopagebreak
  \includegraphics[width=8.5cm]{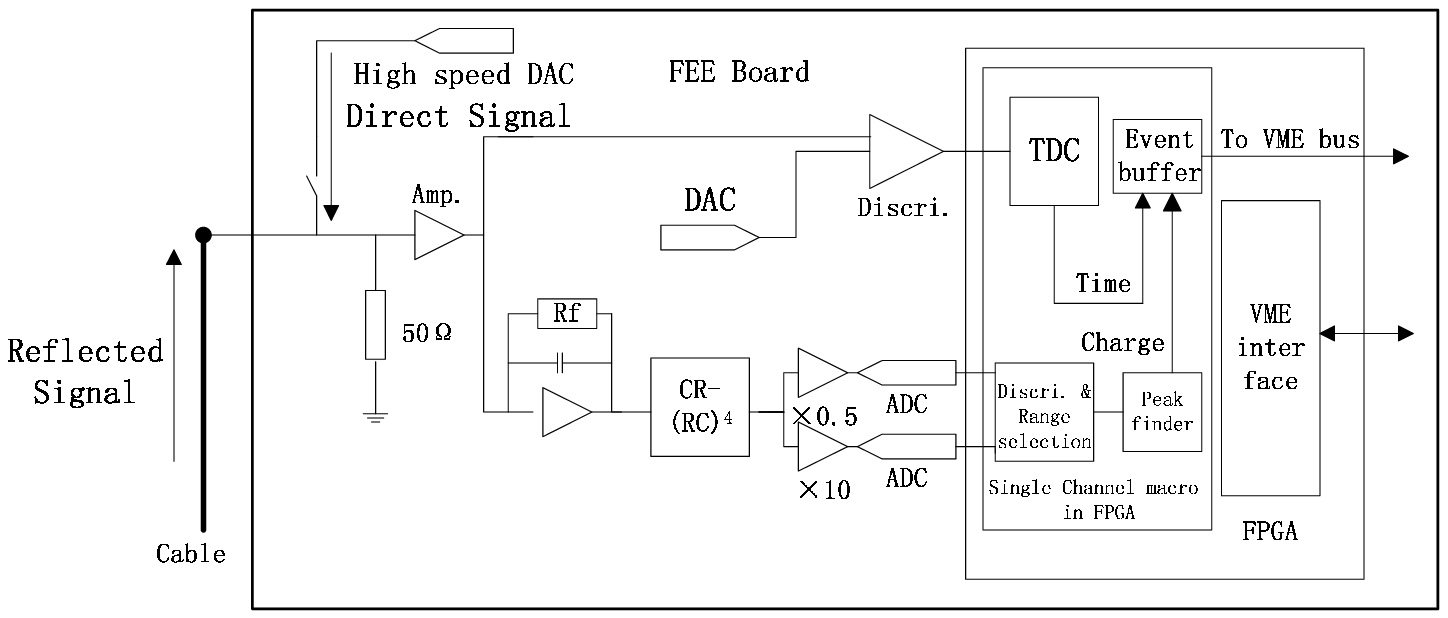}\\

  \figcaption{Schematic diagram of the real-time cable transit time calibration. The pulses generated by the DAC are divided into two. One goes directly to the ADC and TDC inputs. The other is reflected at the connection with the detector and returns to the FEE board.}\label{fig2}
\end{center}
The high speed DAC provides a calibration pulse, with variable voltage, fixed current and output impedance of 50 $\Omega$. The pulse is then divided into two in the FEE board. Part of the signal is sent directly to the TDC and ADC through the electronic circuits. The other part is reflected at the detector due to the mismatched impedance and returns to the FEE board. The input impedance of the PMT in the detector is designed to be large, so that the signal will be totally reflected. In this experiment, the cable type is RG58A/U, one end of which is connected to one channel of the FEE board, while the other end of the cable is open to obtain the high impedance in a simple way.

The time-walk effect of the signal is described by Eq. (\ref{eq2}). The delay caused by the time-walk effect decreases with the increase of the charge $Q$. We believe that the true start time of the pulse is $T=T(\infty)=T_0$ . It is reasonable that with the increase of $Q$, the rise time for the pulse to reach the threshold becomes negligible. $T_{dir0}$ and $T_{ref0}$ are the start times of the direct pulse and reflected pulse respectively, which can be obtained by Eq. (\ref{eq2}). Then, the cable length is
\begin{equation}\label{eq3}
  l=\frac{v\Delta T}{2}=\frac{v(T_{ref0}-T_{dir0})}{2},
\end{equation}
where $\Delta T$ is the time interval between direct and reflected pulses, i.e. the signal transit time through the cable. The pulse signal will be attenuated during the transmission procedure. The propagation velocity and the attenuation vary with the frequency, which results in the distortion and deformation of the shape of the reflected pulse. In this case, the time-walk factors $m$ and $k$ in the Eq. (\ref{eq2}) are estimated  for the direct and reflected pulses respectively.

A typical calibration is shown in Fig. \ref{fig3}. The register of the DAC is set from 500 to 12000 with a step of 1. The measured charge is 0-336 pC for direct pulses and 0-12 pC for reflected pulses. The black dots are the threshold trigger times of the pulses measured at each value of the DAC. The blue blocks are the average charge of the black dots in the same TDC bin. The solid curve is the best fit to those blocks using Eq. (\ref{eq2}). $T_{dir0}$ and $T_{ref0}$ are then calculated.  Here, $m_{dir}$ is 1/2 and $m_{ref}$ is 3/2, as determined by the rise-time shape of the pulses. If the cable length is determined, the shapes of the pulses can be regarded as constant due to the relatively small variation range. The weight of each TDC bin in the calibration, caused by the selection of the DAC values, will influence the validity of the time-walk model. For example, if the pulses all have a large charge, the times measured will be all located in one TDC bin, and then the model of Eq. (\ref{eq2}) does not apply. Therefore, in this paper, the average of the charge in the same TDC bin is used in fitting and the weights of the mean charges are the same. The systematic and statistical errors are estimated in the next section.
\begin{center}
  \includegraphics[scale=0.38]{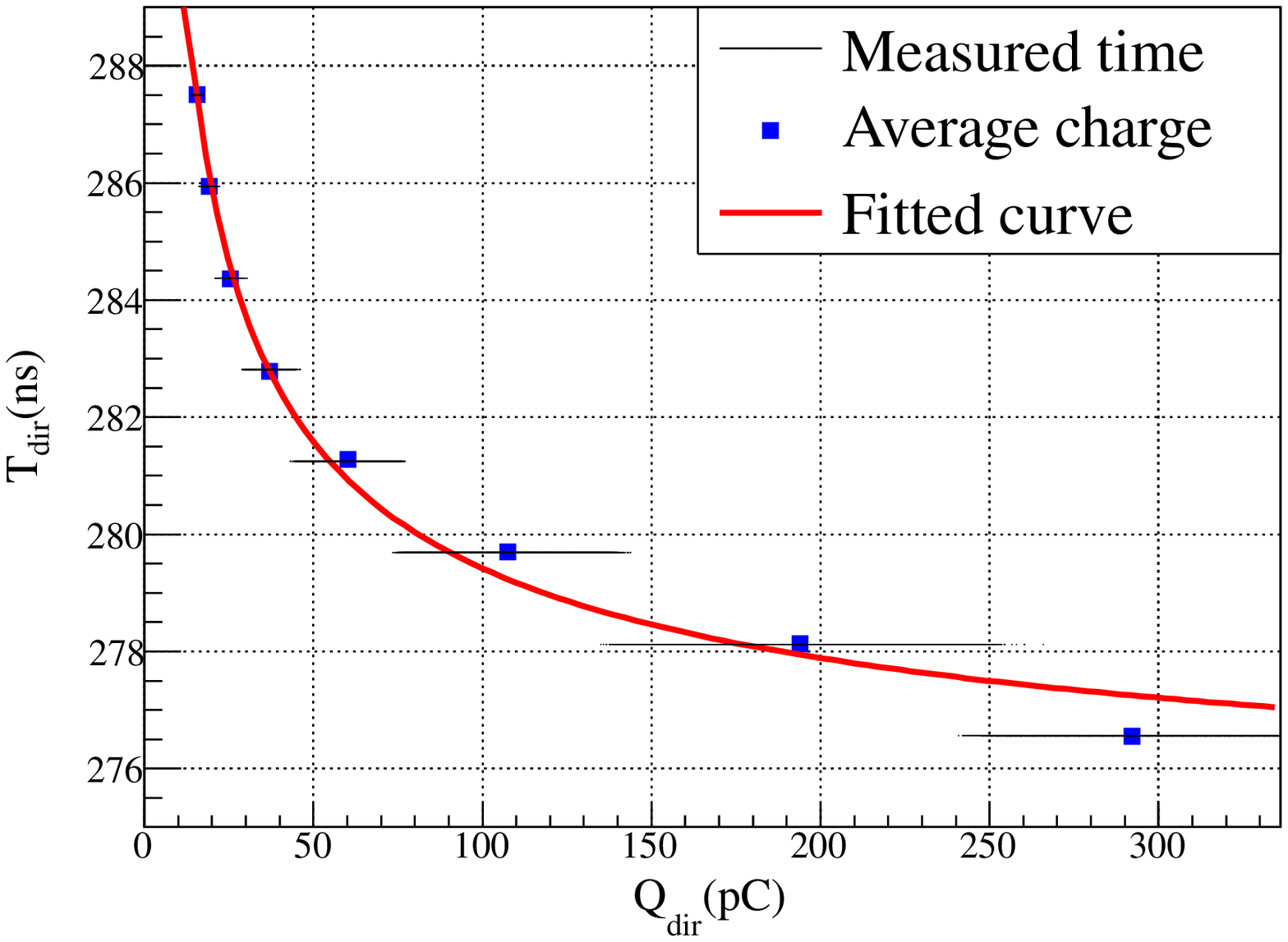}\\
  $(a)$\\
  \includegraphics[scale=0.38]{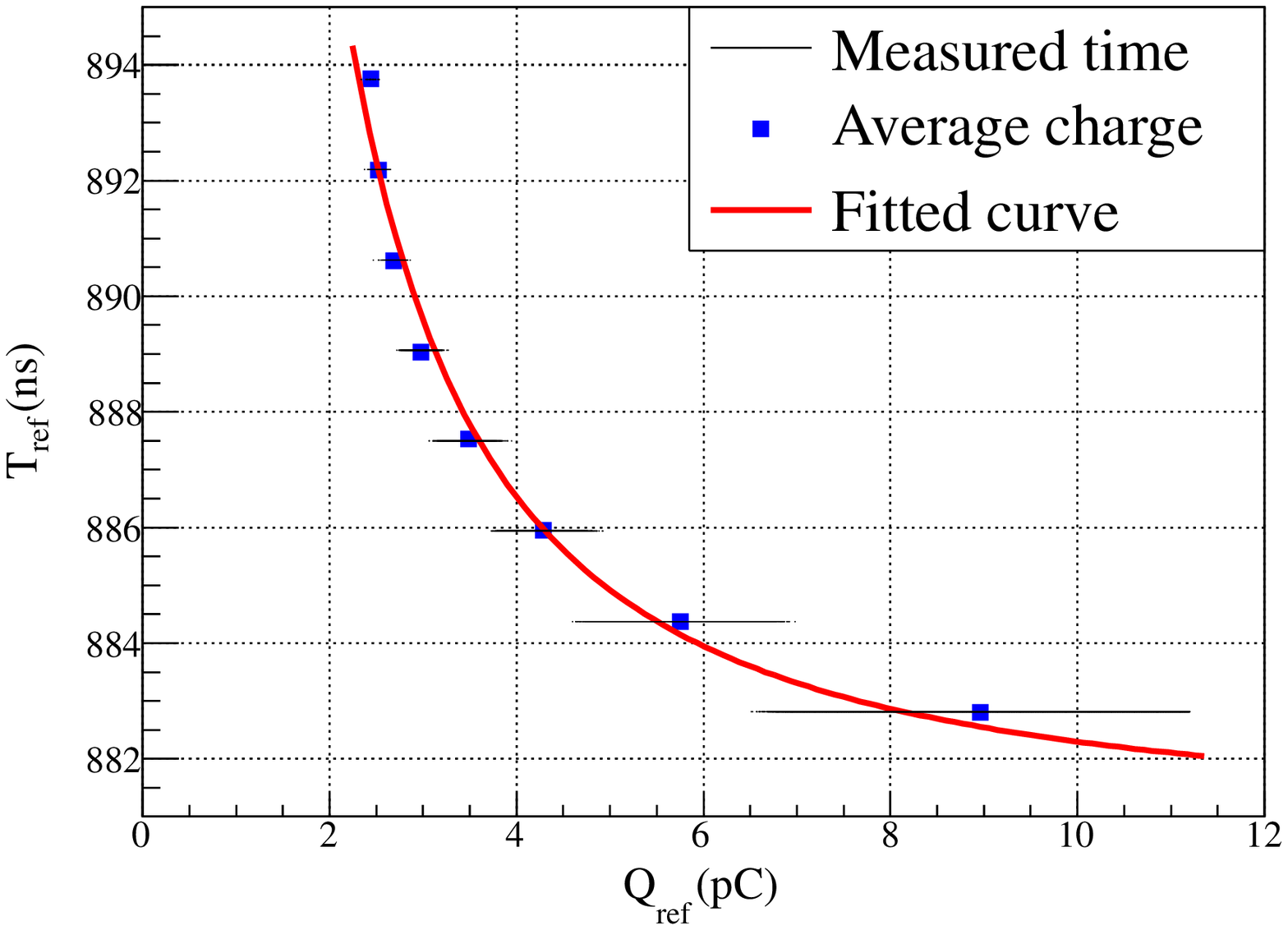}\\
  $(b)$\\
  \figcaption{The measured start time $T_{th}$ as a function of the charge $Q$ for (a) the direct pulse and (b) the reflected pulse. The black dots are the threshold trigger times of the pulses measured at each value of the DAC; the blue blocks are the average charge of the black dots in the same TDC bin; and the solid curve is the best fit to those blocks using Eq. (\ref{eq2}).}\label{fig3}
\end{center}
\section{Calibration precision}

A demonstration experiment was carried out to test the reliability of this real-time calibration method. We used cables with different lengths to simulate the thermal expansion and contraction effect caused by the ambient temperature. Because the cable lengths can be known accurately, it is possible to quantitatively test the calibration method. In order to test small changes in long cables, several short cables of different lengths were connected to the long cable. The incremental lengths of the cable ($dL$) were fixed at 2.75 cm, 5.75 cm, 7.75 cm, 10.65 cm and 13.75 cm.

\begin{figure*}[!t]
  \centering
  \includegraphics[scale=0.8]{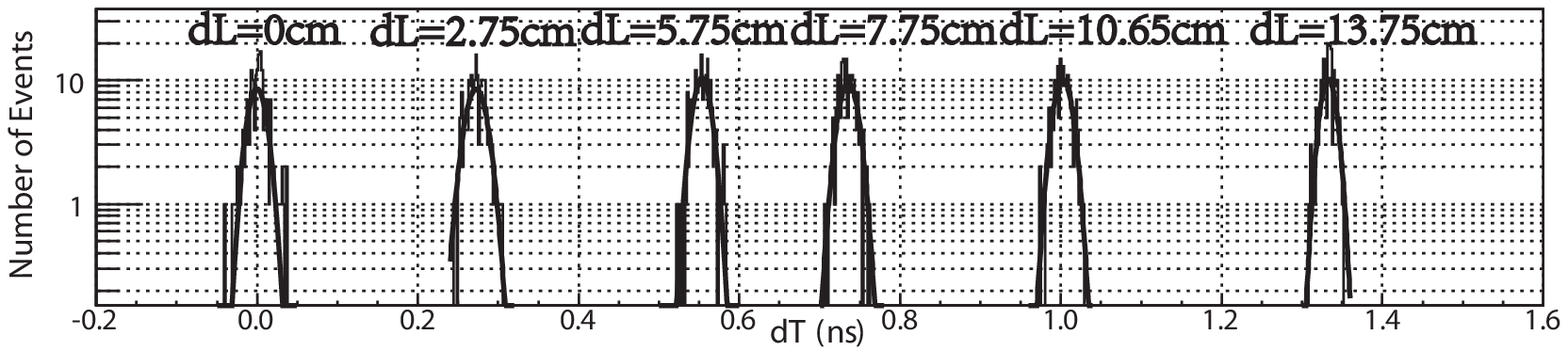}\\
  \caption{Incremental transit time $dT$ measured at different $dL$, which are 2.75 cm, 5.75 cm, 7.75 cm, 10.65 cm and 13.75 cm, respectively. The widths of the Gaussian distributions are $\sim$ 0.01 ns.}\label{fig5}
\end{figure*}

Fig. \ref{fig4} shows the incremental transit time $dT$s for the various values of $dL$. For each $dL$, the calibration was performed about 180 times. Each point is the $dT$ obtained from fitting the direct pulse and reflected pulse as described in the last section. $dT$ increases with $dL$. The value of $dT$ is stable for the same $dL$ as the calibration is repeated. The distribution of $dT$ for each value of $dL$ is obtained and presented in Fig. \ref{fig5}. The distributions obey a Gaussian distribution, and the width of the Gaussian distribution represents the statistical error for each calibration.

\begin{center}
  \includegraphics[scale=0.42]{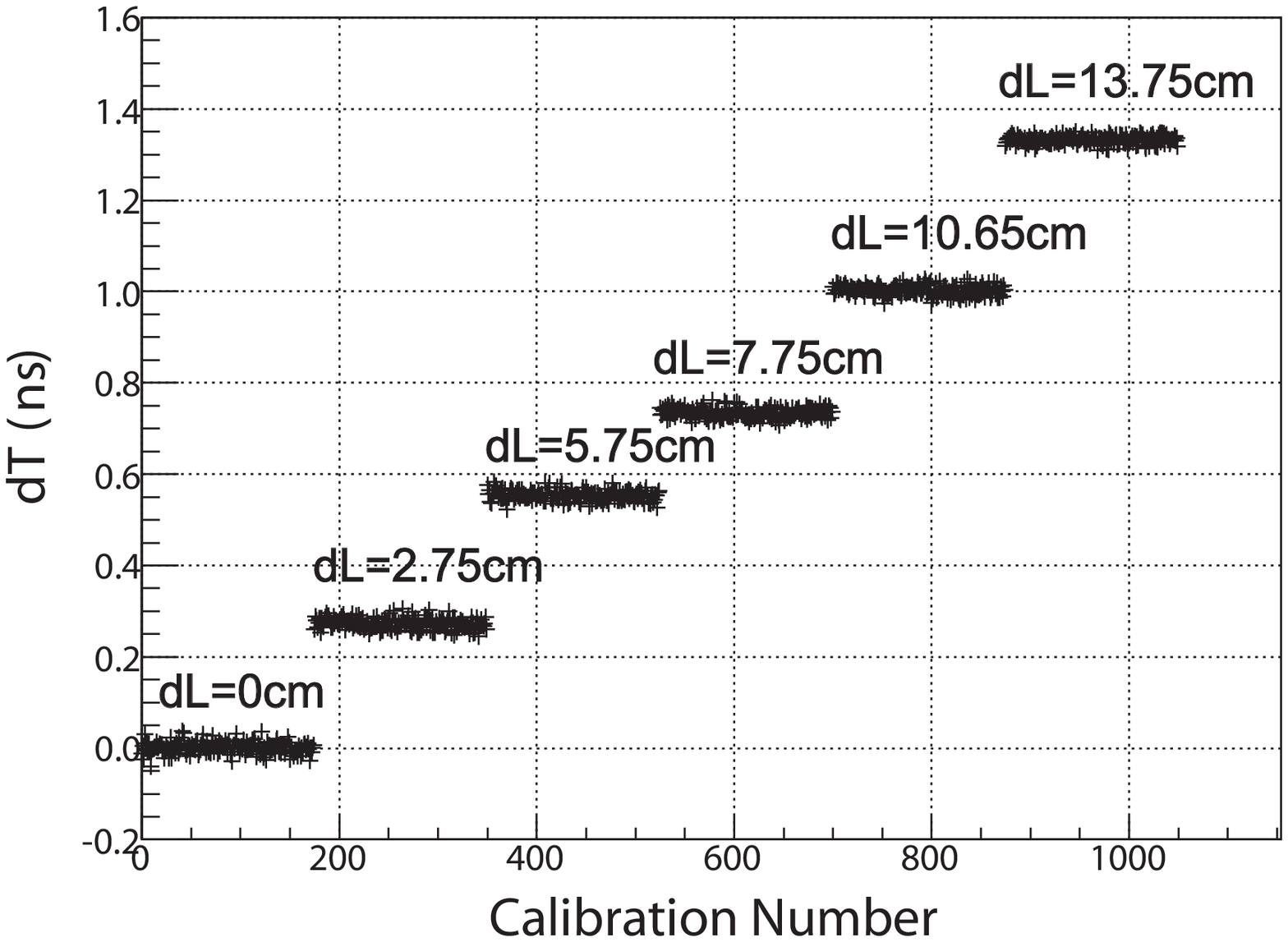}\\
  \figcaption{Incremental transit time $dT$ as a function of the calibration number. $dT$ is measured at different $dL$, which is set at 2.75 cm, 5.75 cm, 7.75 cm, 10.65 cm and 13.75 cm. The cable is calibrated every 15 minutes and repeated $\sim$ 180 times for the same length. }\label{fig4}
\end{center}

The dependence of $dT$ on $dL$ represents the signal propagation velocity in the cable, which is found to be $2.083\pm0.011\times10^8 m/s$ by fitting these values of $dT$ and $dL$. The propagation velocity is related to the dielectric constant and impedance of the cable, and this value agrees with our expectation.

In real-time calibration, the cable is calibrated at regular intervals ($\sim$ 20 minutes). One typical calibration at each $dL$ is shown in Table \ref{tab1}. The cable length was measured using a vernier caliper. The expected $dT$ is the ratio of twice the $dL$ to the propagation velocity. The statistical error of the measured $dT$ is estimated from Fig. \ref{fig5}. As shown in Table \ref{tab1}, the difference between the measured value and expected value is within the statistical error. For the direct pulse and reflected pulse generated from the DAC, the model given by Eq. (\ref{eq2}) can still be optimized, and a better model will give a higher precision. However, the current result is already far better than the precision of 0.1 ns required by the experiment.

\begin{center}
\tabcaption{ \label{tab1}  Incremental transit time $dT$s at various $dL$s for a typical calibration.}
\begin{tabular*}{80mm}{c@{\extracolsep{\fill}}ccc}
\toprule $dL$(mm)  & Expected $dT$(ns) & Measured $dT$(ns) \\
\hline
0                & 0                & $-0.005 \pm 0.011$ \\
27.5 $\pm$ 0.1   &0.264 $\pm$ 0.001 & 0.279 $\pm$ 0.013  \\
57.5 $\pm$ 0.1   &0.553 $\pm$ 0.001 & 0.548 $\pm$ 0.011  \\
77.5 $\pm$ 0.1   &0.745 $\pm$ 0.001 & 0.736 $\pm$ 0.011  \\
106.5 $\pm$ 0.1  &1.025 $\pm$ 0.001 & 1.000 $\pm$ 0.011  \\
137.5 $\pm$ 0.1  &1.323 $\pm$ 0.001 & 1.337 $\pm$ 0.009  \\
\bottomrule
\end{tabular*}
\end{center}

\section{Conclusions and Discussion}
The time domain pulse reflection method allows us to obtain the signal transit time through a cable by measuring the delay time of the direct pulse and the reflected pulse. By changing the amplitude of the calibrating pulse, the dependence of the measured charge and start time can be described by the time-walk model. Using this model, the measurement of the transit time through a cable is improved, giving a high precision. Using the time-walk effect model, the time domain pulse reflection method and the Daya Bay FEE board, we have successfully achieved a precision of better than 0.1 ns even when there is poor TDC resolution.

In principle, the statistical error could be largely reduced by increasing the number of calibrations. However, the systematic error will be dominant, because the simple time-walk model does not describe the experimental facts accurately. In EAS experiments, the transit time of the signal through the long cable is about one microsecond, with a maximum variation of a few nanoseconds caused by the ambient temperature. Therefore, as long as the electronics are reliable, this model can fully satisfy our calibration needs. Furthermore, the waveform of the calibration signal and the time-walk model will be optimized to give an even better precision in the near future.
\subsection*{Acknowledgments}
\acknowledgments{This work was supported by the Ministry of Science and Technology of China, the Natural Sciences Foundation of China (Nos. 11135010, 11105156), by the 973 Program of China(No. 2013CB837000), and by the Youth Innovation Promotion Association, Chinese Academy of Sciences.}
\end{multicols}

\vspace{-1mm}
\centerline{\rule{80mm}{0.1pt}}
\vspace{2mm}

\begin{multicols}{2}

\end{multicols}

\end{CJK*}
\end{document}